\documentclass[aps,prc, groupedaddress,superscriptaddress,twocolumn,showpacs]{revtex4}
\usepackage{graphicx}
\usepackage{longtable}
\usepackage{array}
\begin {document}
\author {S.~S.~Ratkevich}
\author {I.~D.~Fedorets}
\affiliation{Karazin Kharkiv National University, Ukraine}
\author {B.~A.~Nemashkalo}
\affiliation{Kharkov Institute for Physics and Technology, Ukraine}
\author {V.~E.~Storizhko}
\affiliation{Institute for Applied Physics, NASU, Sumy, Ukraine}


\title{Radiative Strength Functions for Dipole Transitions in $^{57,59}$Co}

\begin{abstract}
Average-resonance data on the $(p,\gamma)$ reaction on $^{56,58}$Fe that were taken at proton energies of $E_p=1.5\div3.0$ MeV are used to determine the absolute  values of the radiative strength functions for energies below 10 MeV.
The results obtained in this way are compared with the results of the calculations that rely on the statistical approach and which take into account the temperature of the nucleus and its shell structure. Good agreement with experimental data is achieved without any variation of parameters.
\end{abstract}

\maketitle

 \section{Introduction}
 Experimental and theoretical investigations of the energy dependence of radiative strength functions for nuclei whose shells are filled almost completely or completely revealed that, if the Lorentz distributions that describe well the electric giant dipole resonances excited in such nuclei are extrapolated to the region of low energies, the resulting curve complies with experimental data neither in absolute value nor in shape. For example, an extrapolation of a Lorentzian curve to the region of low energies of gamma rays yields radiative-strength-function values that are eight times as great as the corresponding experimental values obtained in [1] for the $^{59}$Co nucleus, which has a nearly filled proton shell $(Z=27)$. Attempts undertaken in [1] to change the absolute values of the radiative strength function by varying parameters used in determining this function proved to be futile; therefore, its behavior was considered to be anomalous. On the other hand, the same authors [2] obtained data on the radiative strength function for the $^{65}$Cu nucleus that were in good agreement with the extrapolation of the corresponding Lorentzian form. A deviation of the radiative strength function from the Lorentzian behavior is at odds with the well-known Brink hypothesis. In accordance with this hypothesis, primary $E1$ transitions that are observed in radiative nucleon capture are associated with the same processes as giant dipole resonances approximated by a Lorentzian form; moreover, giant resonances built on the ground state and on excited states of the final nucleus are described in terms of the same parameters. The above deviations may suggest the nuclear-structure dependence of the radiative strength function.

The objective of the present study is to determine the absolute values of the radiative strength functions
for the electric dipole transitions in $^{57,59}$Co nuclei near the nucleon binding energy and to analyze their energy dependence. We determine here the relevant radiative strength functions from the averaged intensities of primary gamma transitions that proceed to individual low-lying states of the nuclei being investigated and which are excited in the $(p,\gamma)$ reactions on $^{56,58}$Fe nuclei at incident-proton energies between 1.5 and 3.0~MeV. The energy $Q$ of the $(p,\gamma_0)$ reactions on these target nuclei is 6.02~MeV for  $^{56}$Fe and 7.37~MeV for  $^{58}$Fe. These values of $Q$ are sufficiently large for the densities of states in compound nuclei to satisfy the requirements that ensure the applicability of the statistical description. The thresholds for the $(p,\gamma n)$ reactions on  $^{56}$Fe and  $^{58}$Fe nuclei exceed 5 and 3~MeV, respectively. Owing to this, investigations could be performed over a wide range of incident-proton energies below the neutron threshold.

\section{Experimental results and their analysis}
Following [1], we determined the radiative strength functions in question by the method of averaging over the resonances of a compound nucleus formed upon incident-proton capture by the target nucleus. This averaging, which is necessary for effectively suppressing Porter-Thomas fluctuations [3] and for achieving a satisfactory statistical accuracy, was ensured by an optimal choice of target thicknesses and by a successive addition of gamma-ray spectra measured at different energies with a step equivalent to the target thickness. In taking an average over an interval of width 180~keV for $^{57}$Co and an average over an interval of width 220~keV for $^{59}$Co, the scatter of data that is associated with Porter-Thomas fluctuations did not exceed the statistical uncertainty of measurements, which was within 20\%.

\begin{figure}[ht]
\begin{center}
\includegraphics*[width=6.7cm,angle=270.]{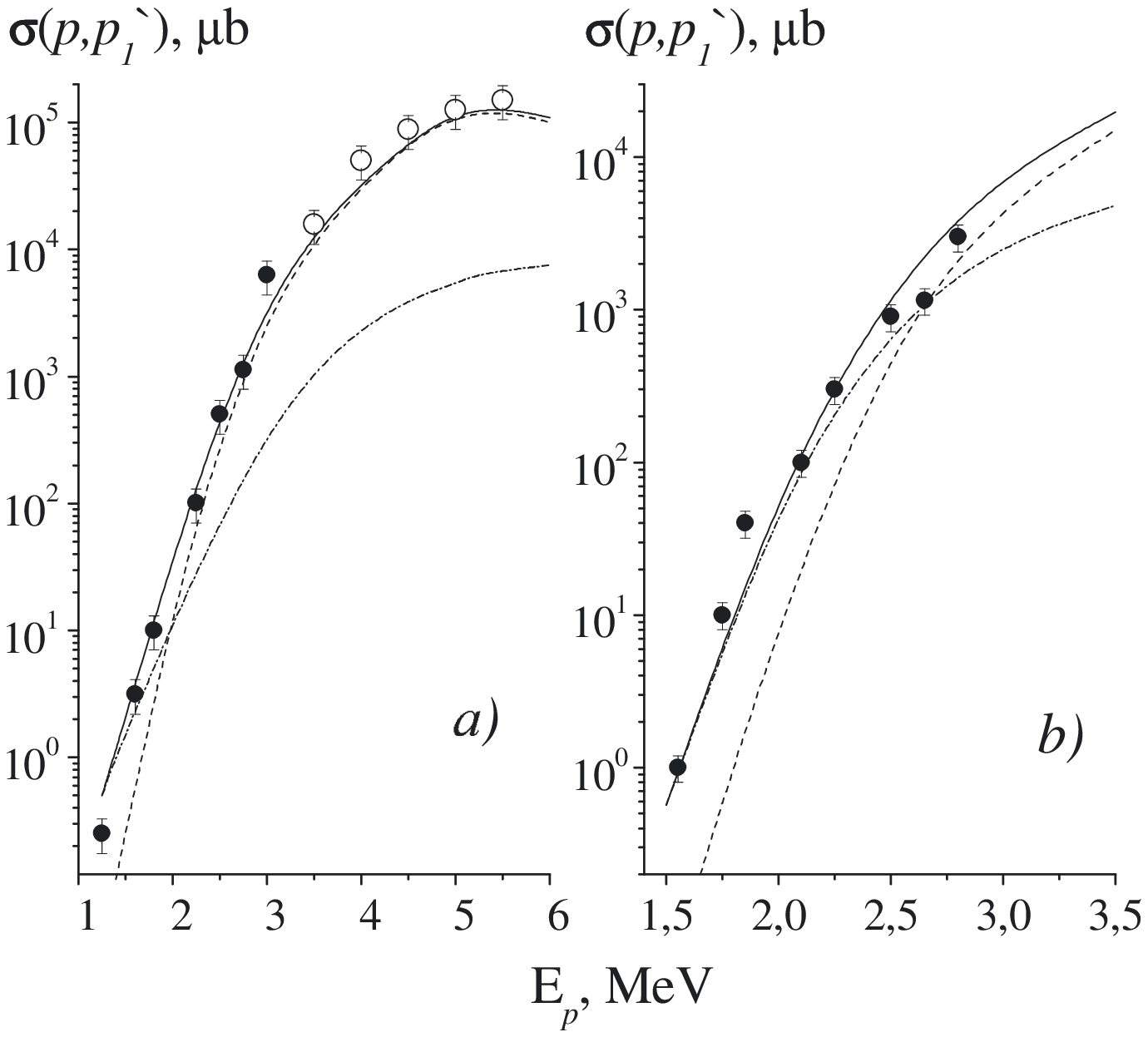}
\caption{Cross sections for inelastic proton scattering on ({\it {a}}) $^{56}$Fe and ($b$) $^{58}$Fe target nuclei (experimental points and calculated
curves): (closed and open circles) experimental data from [11] and [12], respectively; (dash-dotted and dashed curves) contributions
to the calculated cross sections from, respectively, Coulomb and nuclear scattering; and (solid curve) the sum of the Coulomb and
nuclear contributions.}
\label{pic1}
\end{center}
\end{figure}

We used targets manufactured by electrolytically precipitating, onto a gold substrate, $^{56}$Fe (the degree of enrichment was 99.9\%) in order to obtain 849~$m$g/cm$^2$-and 1.729~$m$g/cm$^2$-thick samples or $^{58}$Fe (the degree of enrichment was 90.7\%) in order to obtain 849~$m$g/cm$^2$ - thick samples. The measurements were performed by using protons accelerated by an electrostatic accelerator to energies in the range $1.5\div3.0$ MeV, which was scanned with a variable step equal to proton-energy losses in the target. The spectra of gamma rays corresponding
to primary transitions were measured by a pair spectrometer arranged at an angle of $55^\circ$ to the proton-beam direction. The yields of gamma rays corresponding to direct transitions to the ground states of $^{57}$Co and $^{59}$Co were also measured with the aid of a NaI(Tl) detector of  dimensions $200\times200$~mm$^2$. The strategy of our experiment and the procedure that we used for specific measurements were described in detail
elsewhere [4].

As in [1], radiative strength functions were determined here by using the fact that the radiative-strength function $(S_{\lambda f}(E_\gamma)$ for electric dipole transitions can be related to the proton-capture cross section  $\sigma(p,\gamma_f)$. With the aid of the Hauser–Feshbach formula, the relation in
question can be reduced to the form
\begin{eqnarray}
& &\mathop{\overline {\sigma}_{p,\gamma_f}}=
\frac{{\pi
\mathchar'26\mkern-10mu\lambda _p^2 }}
{{2\left({2I+1}\right)}}
\sum \limits_{\mathop I_{c} \mathop j_{p}
\mathop l_p} {(2I_c + 1)}
\times  \\
\nonumber
& &\times \frac {\mathop T\nolimits_{\mathop l\nolimits_p \mathop j\nolimits_p }\times \mathop {2\pi E}\nolimits_\gamma^3 \mathop S\nolimits_{\lambda f} \left( {\mathop E\nolimits_\gamma  } \right)}
{\sum \limits_{\mathop j_{p'} \mathop l_{p'}}\mathop T\nolimits_{\mathop l\nolimits_{p'} \mathop j\nolimits_{p'}}+\sum \limits_{j'}
\int\limits_0^{\mathop E\nolimits_\lambda}
2\pi\rho_{j'}
(E_\lambda-E_\gamma)
E_\gamma^3 S_{\lambda f}
(E_\gamma)dE_\gamma}
\label{eq1}
\end{eqnarray}
where $\lambda_p$ is the incident-proton wavelength; $I$ is the spin of the target nucleus; $I_c$ is the spin of the compound nucleus; $j_p$ and $\l_p$ are, respectively, the spin and the orbital angular momentum in the input channel; $j_{p'}$ and $l_{p'}$ are the corresponding quantities in the output channel involving proton emission; $S_{\lambda f}(E_\gamma)=S_{\lambda f}^{E1}(E_\gamma)+S_{\lambda f}^{M1}(E_\gamma)$ is the sum of $E1$ and $M1$ radiative strength functions for transitions from the group $\lambda$ of com-pound-nucleus states at energy $E_\lambda$ to the state of energy $E_f$; $T_{l_pj_p}$  and $T_{l_{p'}j_{p'}}$ are the penetrability factors for protons in the input and the output channel, respectively; and $\rho_{j'}(E_\lambda-E_\gamma)$ is the density of levels characterized by a spin $j'$ and an excitation energy $E=(E_\lambda-E_\gamma)$. In our calculations, we took into account the correction for cross-section fluctuations of the Ericson type, which arise because of a small number of open channels, since such a correction may prove to be of importance at low energies [5]. It was assumed in [1] that, for $E1$ transitions, the dependence of the radiative strength function on the energy $E_\gamma$ has the form
\begin{eqnarray}
S_{\lambda f}(E_\gamma)=a \frac{10^{-14}}{2\pi}A^{8/3}E_\gamma^{k-3}{\rm{(MeV^{-3})}},
\label{eq2}
\end{eqnarray}
where $a$ and $k$ are parameters, whose values are fixed in fitting expression (2) to experimental data. The value of $k=4.7$ was obtained in [1] by extrapolating the Lorentzian form that describes the giant dipole resonance in $^{59}$Co to the energy region under study. For the case of a direct $\gamma_0$ transition to the ground state of the $^{59}$Co nucleus, a least squares fit of the theoretical cross section $\sigma(p,\gamma_0)$ to its experimental value yielded $a=1.5$ [1]. The radiative strength function as determined by using the above values of the parameters $a$ and $k$ reproduces the slope of the Lorentzian curve, but the absolute values of this function differ from that which would be expected on the basis of extrapolation by a factor of 8 [1].

\begin{figure}[ht*]
\begin{center}
\includegraphics*[width=5.8cm,angle=270.]{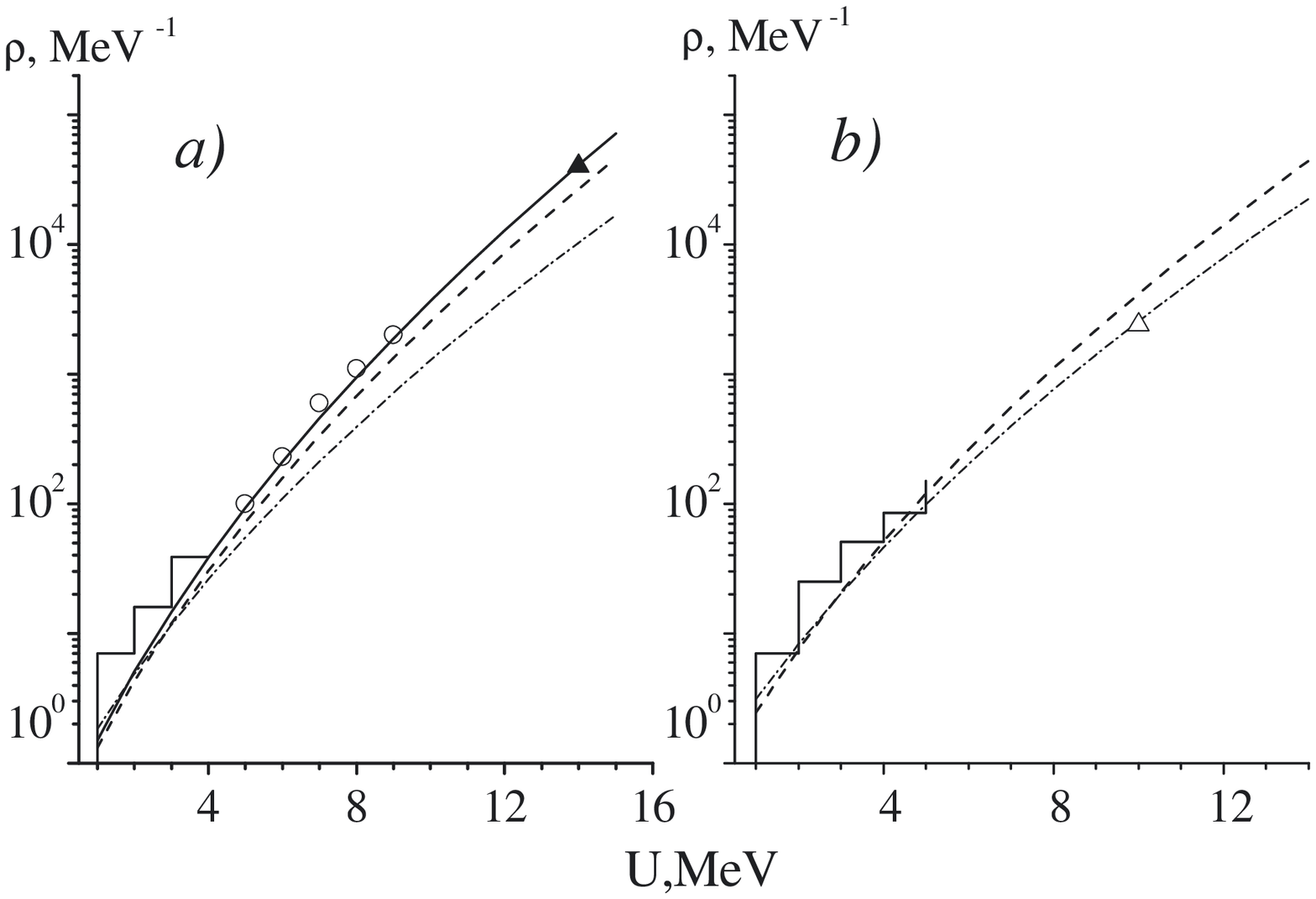}
\caption{ {\small {Level densities in the ($a$) $^{57}$Co and ($b$) $^{59}$Co nuclei
versus excitation energy: (broken lines) discrete levels
established in [14, 15], (open circles) experimental data
from [17], (closed triangle) experimental result from [18],
(open triangle) experimental result from [19], (dash-dotted
curve) results of the calculation within the back-shifted
Fermi gas model with the parameter values from [20] and
the nuclear moment of inertia set to half the rigid-body
value, (dashed curve) results of the calculation within the
back-shifted Fermi gas model with the parameter values
from [20] and the nuclear moment of inertia set to the rigidbody
value, and (solid curve) results of the calculation with
the parameter values adopted in the present study.}}}
\label{pic2}
\end{center}
\end{figure}

In the present study, the radiative strength function appearing in the denominator on the right-hand side of (1) was parameterized either in a Lorentzian form or in that form which was obtained in the approach developed in [6,7] on the basis of Fermi liquid theory. The quantity $S_{\lambda f}(E_\gamma)$ in the numerator was chosen in such a way as to reproduce the absolute values of the partial cross section that were obtained experimentally. The penetrability factors for protons were calculated with allowance for the results reported in [8-10]. The parameters of the optical potential were determined from the best fit to the experimental cross section for the reactions $^{56,58}$Fe$(p,p'\gamma)$ (Fig.1), $^{56,58}$Fe$(p,\gamma)$, and $^{58}$Fe$(p,n)$ in the region of incident-proton energies below 4~MeV. The experimental cross-section values presented in Fig.1 were borrowed from [11,12] for inelastic proton scattering on $^{56}$Fe nuclei and from [13] for inelastic proton scattering on $^{58}$Fe nuclei. For the geometric parameters of the real part of the optical potential, we chose the values \\
\begin{center}
$r_r=1.17 \mathop {}\nolimits_{}^{}{\rm{fm,}}
\mathop {}\nolimits_{}^{}\mathop {}\nolimits_{}^{}
r_s=1.32
\mathop {}\nolimits_{}^{} \rm{fm,} $
\end{center}
\begin{center}
$a_r=0.70 \mathop {}\nolimits_{}^{}{\rm{fm}},
\mathop {}\nolimits_{}^{}\mathop {}\nolimits_{}^{}
a_s=0.58 \mathop {}\nolimits_{}^{}{\rm{fm}}. $
\end{center}

\begin{table}
\caption{$(p,\gamma_f)$ partial cross sections for $^{56}$Fe targets}
\begin{tabular}{c|c|c|c|c|c}

\hline
            &          &                        &                      &        &   \\
            &          & $\sigma^{exp}$,~$\mu$b&   $\sigma_{1}^{calc}$,& $\sigma_{2}^{calc}$, &$\sigma_{3}^{calc}$,\\
$E_{f}$,~MeV & $J^{\pi}$&   $E_{ð}=2.84$MeV     &   $\mu$b &$\mu$b &$\mu$b  \\
            &         &$ \Delta Å_{ð}=180$~keV  &          &       &         \\ \hline\hline
 0.0        & 7/2$^{-}$&   23.0$\pm4.4$         &    37.53 & 19.57 &   26.55 \\ 
 1.224      & 9/2 $^{-}$&   5.6$\pm1.5$         &    3.44  &  2.96 &   4.50\\ 
 1.378      & 3/2 $^{-}$&  24.9$\pm4.7$         &   50.63  & 14.74 &  23.15\\ 
 1.505      & 1/2 $^{-}$&  16.4$\pm6.3$         &   33.42  &  8.97 &  13.58\\ 
 1.690      &11/2 $^{-}$&   1.1$\pm0.5$         &   1.56   &  0.93 &   1.21\\ 
 1.758      & 3/2 $^{-}$&  18.5$\pm4.2$         &   40.15  & 11.16 & 16.00 \\ 
 1.897      & 7/2 $^{-}$&   7.2$\pm1.8$         &   12.86  &  5.22 &   6.55\\ 
 1.920      & 5/2 $^{-}$&  11.3$\pm3.1$         &   19.49  &  6.77 &  8.79 \\ 
 2.133      & 5/2 $^{+}$&  14.4$\pm4.4$         &   38.32  &  8.43 & 13.36 \\
 2.133      & 5/2 $^{-}$&                       &          &       &       \\ 
 2.479      & 3/2$^{-*}$&   4.9$\pm2.0$         & 18.02    &  5.52 &   5.85\\
 2.485      & 9/2 $^{-}$&                       &          &       &       \\ 
 2.514      & 7/2$^{-*}$&               &&&                                \\
 2.523      & 13/2$^{-}$  & 3.0$\pm1.5$         &   19.02  &  4.15 &   4.51 \\
 2.560      & 9/2 $^{-}$  &              &&&                                \\ 
 2.611      & 7/2 $^{-}$  &     3.3$\pm1.6$ &24.54   & 6.69  &  7.31    \\
 2.615      & 9/2 $^{-*}$ &              &&&                            \\  
 2.723      & 9/2 $^{-*}$ &              &&&                            \\
 2.731      & 3/2 $^{-}$  &   12.0$\pm4.1$  &30.1    &7.13   &9.89      \\
 2.743      & 11/2 $^{-}$ &              &&&                           \\ 
 2.804      & 5/2 $^{-}$  &   9.2$\pm3.5$ &30.20   & 8.41  & 8.65       \\
 2.879      & 3/2 $^{-}$  &              &&&                            \\ 
 2.981      & 1/2 $^{+}$  &    8.0$\pm3.0$ &22.36  &  5.05 &  8.63      \\
 2.982      & 5/2 $^{-*}$ &              &&&                             \\ \hline

\end{tabular}
\footnotetext[1]
 {\small{{\it {The superscripts "exp" and "calc" label, respectively, the experimental and calculated cross-section values; the subscripts {"\rm{1}"}, {"\rm{2}"}, and {"\rm{3}"} on the latter label the theoretical values corresponding, respectively, to the radiative strength functions in a Lorentzian form,
to the radiative strength functions calculated with allowance for only the nuclear temperature, and to the radiative strength functions
calculated with allowance for nuclear temperature and shell effects. Asterisks indicate spin–parity assignments chosen in the present
study.}}}}
\end{table}

\begin{table}
\caption{$(p,\gamma_f)$ partial cross sections for $^{58}$Fe targets}
\begin{tabular}{c|c|c|c|c|c}

\hline
            &          & $\sigma^{exp}$,~$\mu$b&   $\sigma_{1}^{calc}$,& $\sigma_{2}^{calc}$, &$\sigma_{3}^{calc}$,\\
$E_{f}$,~MeV & $J^{\pi}$&   $E_{ð}=2.8$MeV     &   $\mu$b &$\mu$b &$\mu$b  \\
            &         &$ \Delta Å_{ð}=220$~keV  &          &       &         \\ \hline\hline

  0.0 & 7/2$^{-}$   &     49.0$\pm3.9$& 110.80 & 51.38 & 52.15\\ 
1.099 & 3/2$^{-}$   &     68.6$\pm6.4$& 170.62 & 57.71 & 74.10\\ 
1.190 & 9/2$^{-}$   &      8.5$\pm5.0$&  17.72 &  6.96 &  7.72\\ 
1.292 & 3/2$^{-}$   &     64.3$\pm9.7$& 153.47 & 51.65 & 68.93\\ 

1.434 & 1/2$^{-}$   &             &        &       &          \\
1.460 &11/2$^{-}$   &     65.8$\pm15$ &  172.14&  61.78 & 68.74\\
1.482 & 5/2$^{-}$   &             &        &       &          \\ 

1.745 & 7/2$^{-}$   &     19.5$\pm4.3$&   43.93&  16.33& 23.73\\ 

2.062 & 7/2$^{-}$   &             &        &       &          \\
2.087 & 5/2$^{-}$   &     29.5$\pm8.3$&   87.56&  32.41& 46.41\\ 

2.154 & 9/2$^{-}$   &      9.5$\pm2.3$&   12.17&   6.42&  8.03\\ 
2.184 &11/2$^{-}$   &             &        &       &          \\

2.205 & 5/2$^{-}$   &     23.7$\pm7.4$&  47.52 & 17.05 & 24.78\\ 

2.395 & 9/2$^{-}$   &      4.4$\pm2.0$&   9.44 &  3.68 &  4.48\\ 

2.479 & 5/2$^{-}$   &     19.3$\pm4.4$&  40.50 &  14.48& 20.32\\ 

2.540 & 5/2$^{-}$   &             &        &       &          \\
2.582 & 3/2$^{-}$   &     68.1$\pm17$ &  123.86&  46.65& 64.52\\
2.586 & 7/2$^{-}$   &             &        &       &          \\ 

2.713 & 1/2$^{-}$   &             &        &       &          \\
2.722 & 9/2$^{-}$   &             &        &       &          \\
2.770 & 3/2$^{-}$   &     99.9$\pm25$      & 148.38   &  58.73 & 89.18\\
2.782 & 5/2$^{-}$   &             &        &       &          \\ 

2.817 & 3/2$^{-}$   &     32.9$\pm8.7$&   86.43&  24.47& 35.45\\
2.826 & 7/2$^{-}$   &             &        &       &          \\ 

2.912 & 3/2$^{-}$   &             &        &       &          \\
2.958 & 5/2$^{-}$   &     30.6$\pm17$ &  147.37&  42.74& 54.93\\
2.966 & 3/2$^{-}$   &             &        &       &          \\ 

3.015 & 7/2$^{-}$   &      4.7$\pm2.2$&   20.35&   7.33&  9.09\\ 

3.063 & 1/2$^{-}$   &             &        &       &          \\
3.082 & 9/2$^{-}$   &     16.7$\pm6.4$&   30.33&  19.78& 21.47\\
3.09  & 7/2$^{-}$   &             &        &       &          \\ 

3.141 & 7/2$^{-}$   &      4.2$\pm1.8$&   19.36&   6.72&  8.06\\ 

3.160 & 3/2$^{-}$   &     23.4$\pm12$ &   56.89&  21.90& 25.50\\ 
3.194 & 5/2$^{-}$   &                 &        &       &       \\

3.220 & 3/2$^{-}$   &     38.4$\pm8.6$&   96.21&  32.46& 43.27\\
3.276 & 3/2$^{-}$   &             &        &       &          \\ 

3.323 & 7/2$^{-}$   &      4.8$\pm1.4$&   17.16&    5.9&  6.80\\ \hline

\end{tabular}
\end{table}

\begin{figure}[ht]
\begin{center}
\includegraphics*[width=5.5cm,angle=270.]{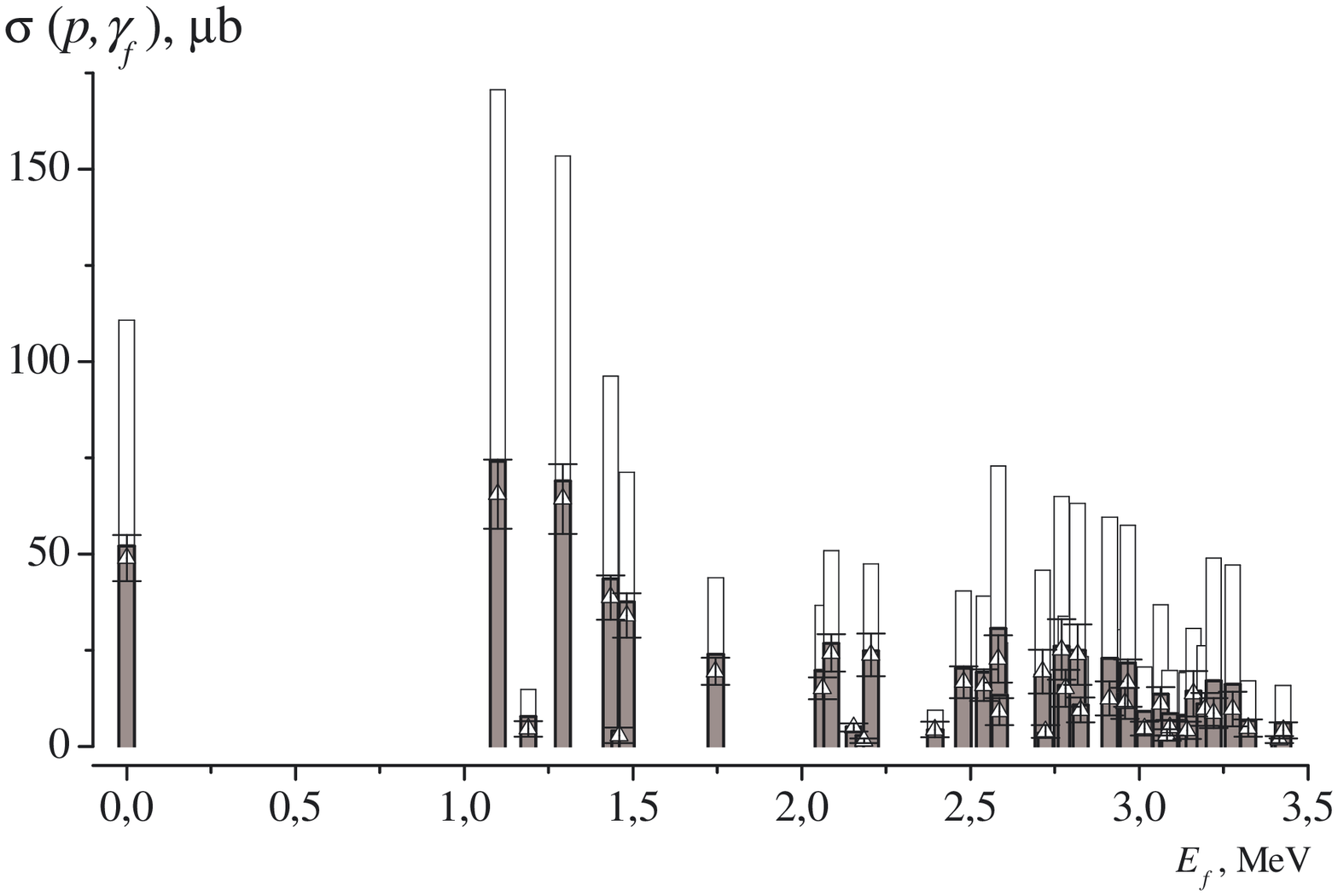}
\caption{$(p,\gamma_f)$ partial cross sections for $^{58}$Fe targets: (shaded areas of the histogram) cross sections calculated with the radiative
strength functions as obtained within the statistical approach [6, 7] allowing for shell effects and for the nuclear temperature,
(unshaded sections of the histogram) cross sections calculated with the radiative strength functions in a Lorentzian form, and (open
triangles) experimental cross sections.}
\label{pic3}
\end{center}
\end{figure}

\begin{figure}[ht*]
\begin{center}
\includegraphics*[width=5.cm,angle=270.]{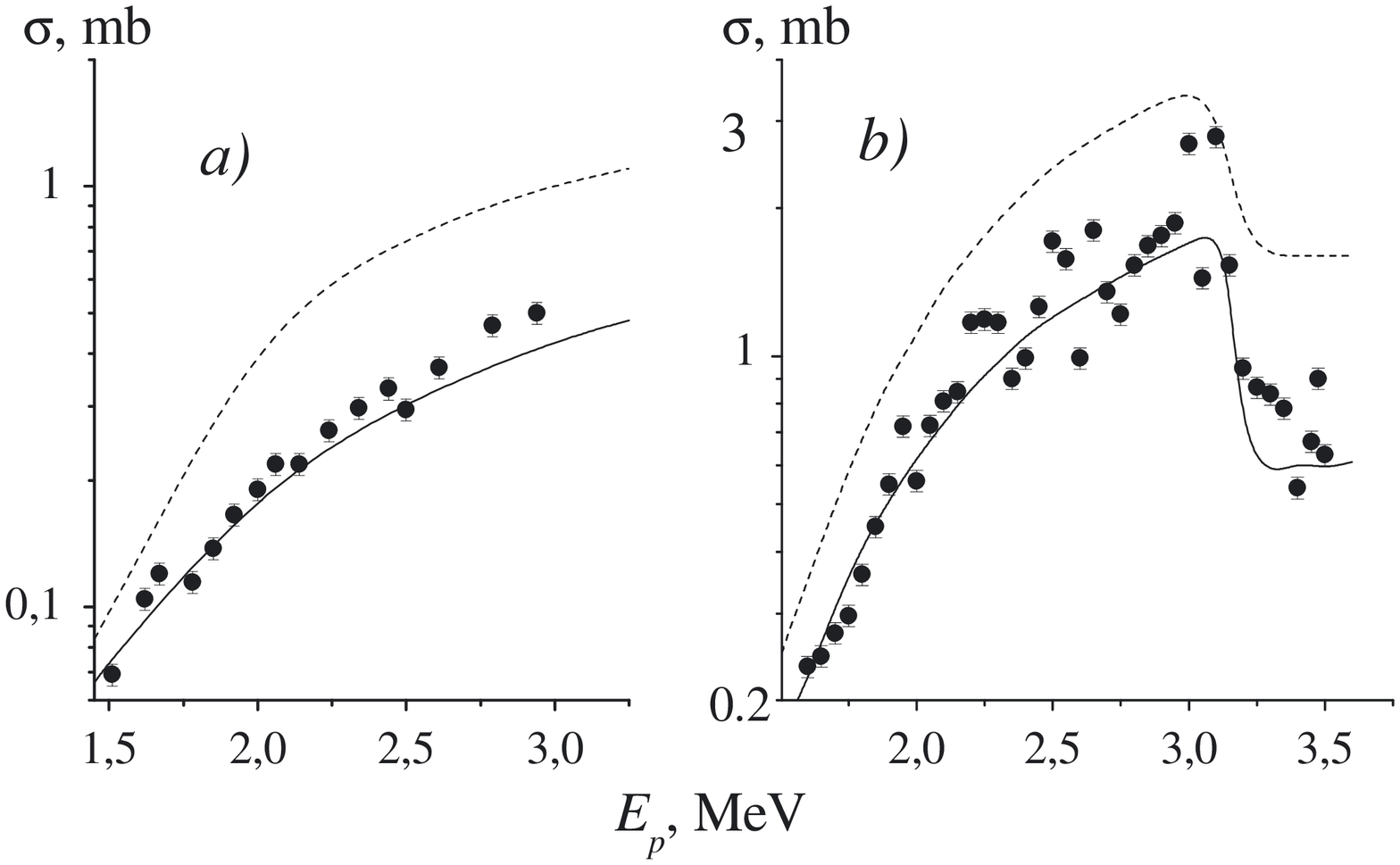}
\caption{ {\small {Total cross sections for the $(p,\gamma)$ reactions on ($a$) $^{56}$Fe
and ($b$) $^{58}$Fe target nuclei: (closed circles in Fig. 4$a$) experimental
cross-section values derived from the estimates of
the results presented in [11, 12], (closed circles in Fig. 4$b$)
experimental cross-section values from [13], (dashed
curves) cross sections computed with the radiative strength
functions in a Lorentzian form, and (solid curves) cross sections
computed with the radiative strength functions as
obtained within the statistical approach developed in [6, 7].}}}
\label{pic4}
\end{center}
\end{figure}

With the exception of the diffuseness parameter set to a value less than that in [8], all the above values are identical to those from that study. For the real part of the potential, we took the values \\
\begin{center}
$ V_r(E)=59.34-0.37E$ for $^{56}$Fe,
\end{center}
\begin{center}
$ V_r(E)=58.0-0.32E$ for $^{58}$Fe,
\end{center}
for the imaginary part of the surface potential, we set
\begin{center}
$W_s(E)=3.85+0.72E$ for $^{56}$Fe,
\end{center}
\begin{center}
$W_s(E)=5.6-0.25E$ for $^{58}$Fe.
\end{center}
In these expressions and in those that precede them, all values are given in MeV.

The parameters of the real part of the optical potential differ only slightly from the global parameter set that is presented in [8] and which was derived on the basis of data on the scattering of protons with energies in excess of 9~MeV; however, the parameters of the imaginary part of the same potential differ from those in the global set more pronouncedly. At the same time, our parameters comply well with the results reported in [10], where an optical-model version that takes into account the dispersion relation between the imaginary and the real part of the potential underlies the description of proton scattering on $^{56}$Fe target nuclei at incident-proton energies between 4.08 and 7.74~MeV.

\begin{figure}[ht]
\begin{center}
\includegraphics*[width=6.2cm,angle=270.]{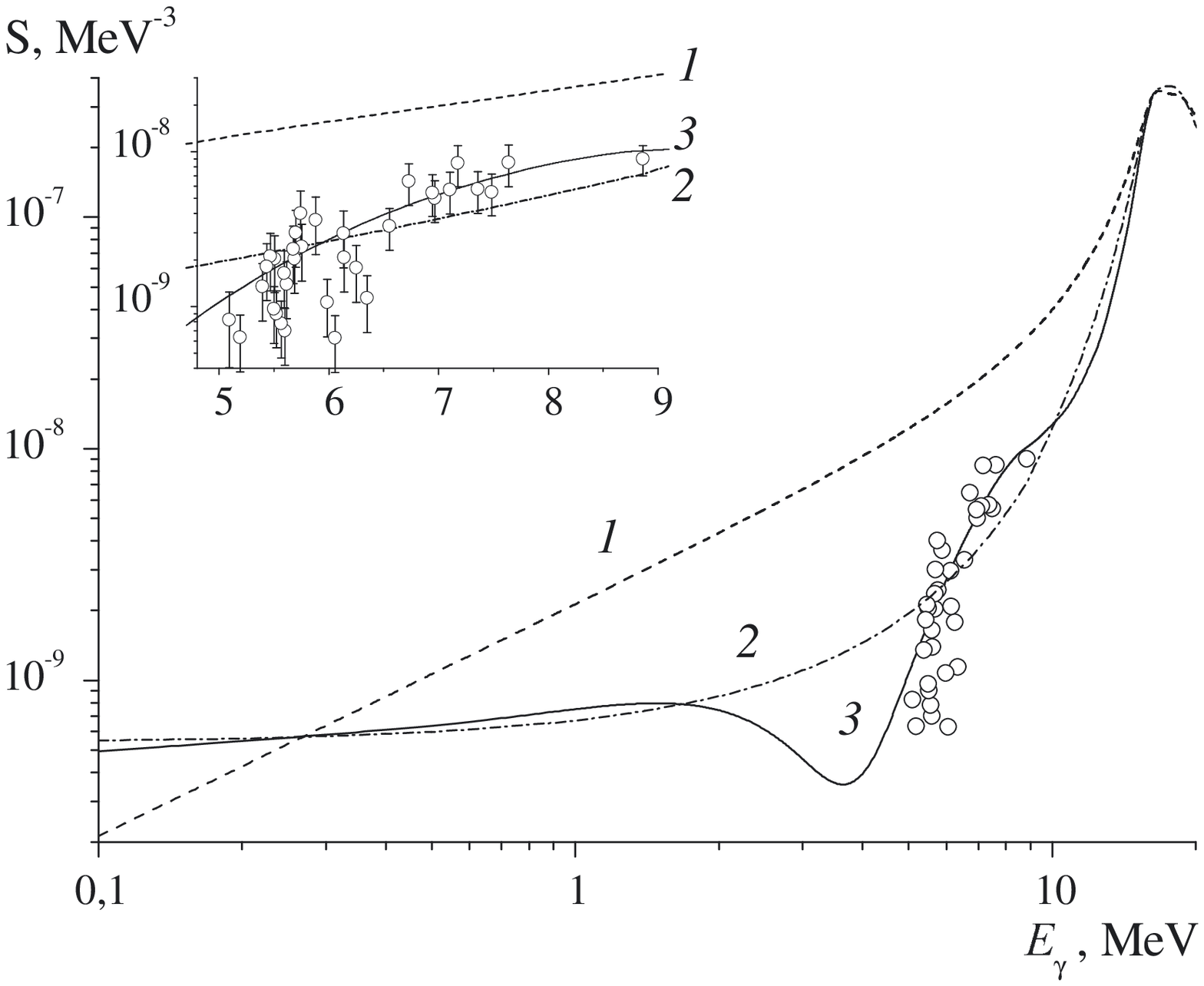}
\caption{ {\small {Experimental and theoretical values of the radiative
strength functions for primary gamma transitions in $^{57}$Co:
(open circles) radiative-strength-function values at $E_p=2.84$ MeV, (curve {\it{1}}) Lorentzian dependence, (curve {\it{2}})
results of the calculations within the statistical approach
with allowance for only nuclear temperature, and (curve {\it{3}})
results of the calculations within the statistical approach
with allowance for the nuclear temperature and shell effects.}}}
\label{pic5}
\end{center}
\end{figure}
\begin{figure}[ht]
\begin{center}
\includegraphics*[width=6.2cm,angle=270.]{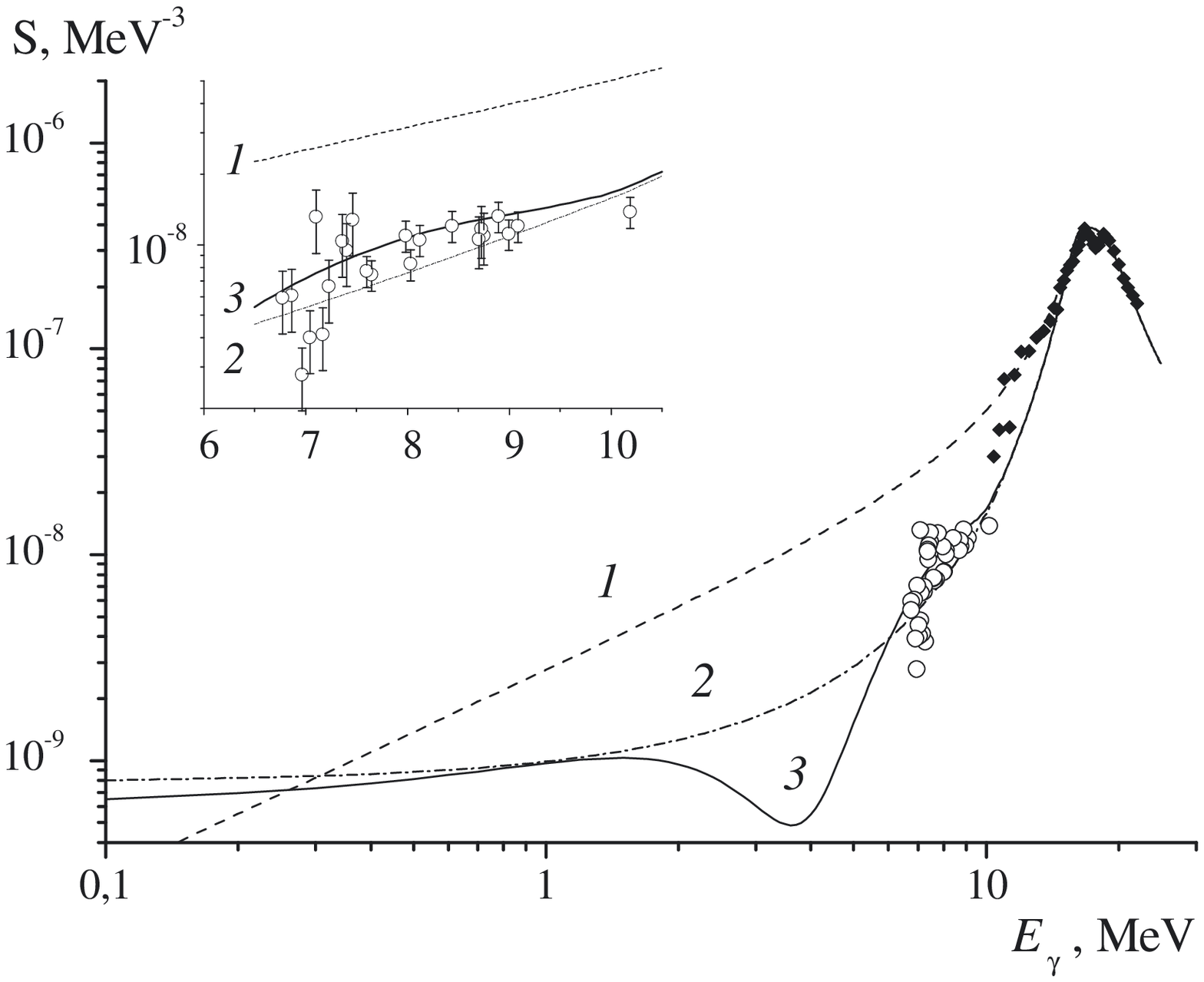}
\caption{ {\small {As in Fig. 5, but for $^{59}$Co. Open circles represent the
radiative-strength-function values at $E_p=2.82$ MeV. Closed
diamonds illustrate data from [21]. The notation for the
curves is identical to that in Fig. 5.}}}
\label{pic6}
\end{center}
\end{figure}

The level densities in the $^{57}$Co and $^{59}$Co nuclei were calculated on the basis of the back-shifted Fermi gas model, with the parameters being set to $a=6.4$ MeV$^{-1}$ and $\Delta=-0.02$ MeV for the former and to $a=5.5$ MeV$^{-1}$ and  $\Delta=-0.77$ MeV for the latter. In these calculations, we used the rigid-body value for the $^{57}$Co moment of inertia and half of it for the $^{59}$Co moment of inertia. These parameter sets ensure the best agreement of the computed values of the level densities (see Fig.2) with data from [14,15] on the discrete section of the energy-level diagram for the nuclei being investigated and with data obtained from an analysis of the experimental spectra of neutrons from $(p,n)$ reactions in the proton energy range $E_p=6\div10$ MeV [16,17], as well as with data deduced from an analysis of Ericson fluctuations  at $U=14$ MeV for $^{57}$Co [18].

The scheme used here to compute radiative strength functions [6] takes into account the dependence of the spread width of the giant dipole resonance on the width of the giant dipole resonance on the
energy $E_\gamma$, the effect of the nuclear temperature, and the
role of shell corrections and of the Pauli exclusion principle.
Within this approach, the $E1$ strength function for the case of a double-peaked giant dipole resonance can
be represented in the form
\begin{eqnarray}
S^{E1}_\gamma & = & 8.674 \cdot 10^{-8}\times
2 \pi\left[1+\exp(-E_{\gamma}/T)\right ]^{-1} \times
\\
\nonumber
&&
 \\
\nonumber
&  &
\times
\sum\limits_{i = 1}^2
\frac{{\sigma_i E_{i}^{2}} \Gamma_{R(i)}(E_\gamma)}
{(E_\gamma^2-E_i^2)^2+E_i
\Gamma_{R(i)}(E_\gamma)}.
 \\
\nonumber
\label{eq3}
\end{eqnarray}
where
\begin{eqnarray}
\Gamma_{R(i)}(E_\gamma)=\Gamma_i\rho_{2p-2h}(E_\gamma,T)/\rho_{2p-2h}(E_i,T).\nonumber
\end{eqnarray}
In these expressions, $\sigma$, $E_i$, and $\Gamma_i$ are, respectively, the
cross sections at the maxima, the positions of the components
of the giant dipole resonance, and their widths.
The values of these parameters are chosen in such a
way as to ensure the best fit of the resulting Lorentzian
shape to experimental data that Alvarez $et$ $al$. [21]
present for the $(\gamma,n)$ reaction on $^{59}$Co. In calculating the
level density $\rho_{2p-2h}(E_\gamma,T)$, we took into account the
shell structure of the spectrum of single-particle levels
and the effect of the nuclear temperature on the occupation
numbers for these nuclei.

In order to calculate the $M1$ strength function, we
made use of the relation [7]
\begin{eqnarray}
S_\gamma^{E1}/S_\gamma^{M1}=0.03A(E^2_\gamma+(\pi T)^2)/B_n^2,
\label{eq4}
\end{eqnarray}
where $B_n$ is the neutron binding energy.

Tables 1 and 2 display the measured values of $(p,\gamma_f)$
partial cross sections for $^{56}$Fe and $^{58}$Fe target nuclei and
the values calculated for these cross sections by formula
(1) with various radiative strength functions. In
order to visualize these results more clearly, the experimental
and the calculated values of the $(p,\gamma_f)$ partial
cross sections for direct gamma transitions to $^{59}$Co
states are shown in Fig. 3 as a histogram. The shaded
areas of the histogram correspond to the cross sections
computed with the radiative strength functions found
within the statistical approach [6,7] with allowance for
the shell structure and nuclear temperature, while the
unshaded areas represent cross sections evaluated with
the radiative strength functions having a Lorentzian
form. In Fig. 4, the theoretical values of the total cross
sections for the $(p,\gamma)$ reactions on $^{56}$Fe and $^{58}$Fe nuclei
are contrasted against relevant experimental data.

The radiative-strength-function values obtained in
the present study from an analysis of the $(p,\gamma)$ reactions
on $^{56}$Fe and $^{58}$Fe nuclei are displayed in Figs. 5 and 6.
The curves in these figures represent theoretical estimates
of the radiative strength function that correspond
to a Lorentzian form (curve {\it{1}}), to the results of the calculations
that are based on expression (3) and which
allow for only temperature (curve {\it{2}}), and to the results
of analogous calculations including both temperature
and shell effects (curve {\it{3}}). The contribution of .1 transitions,
which is not shown in the figures, does not
exceed 15\% for various states of $^{57}$Co and $^{59}$Co. That
the known values of the giant-dipole-resonance parameters
for $^{59}$Co were used throughout for want of experimental
data on the giant dipole resonance in $^{57}$Co obviously
had an adverse effect on the degree of agreement
between the results of our calculations and the experimental
values of the radiative strength functions for
$^{57}$Co.

\section{Conclusion}
Our results indicate that, in the gamma-transition-energy
range under study, the absolute values of the
radiative strength functions for $^{57}$Co and $^{59}$Co fall significantly
short of values on the Lorentzian curves that
describe the corresponding giant dipole resonances. At
the same time, the radiative strength functions as calculated
within the approach developed in [6, 7]—this
approach relies on Fermi liquid theory and takes into
account nuclear temperature and shell effects—agree
with experimental data without the use of adjustable
parameters. This is at odds with the Brink hypothesis,
according to which the radiative strength function for
dipole transitions must not depend on the properties of
the final nuclear state.

\end{document}